\documentclass[twocolumn,letterpaper]{IEEEAerospaceCLS}  

\usepackage{bbm}
\usepackage{verbatim}
\usepackage{cite}
\usepackage{url}
\usepackage[cmex10]{amsmath}
\usepackage{amssymb}
\usepackage{amsmath}
\usepackage[]{graphicx}    
\usepackage{calc}
\usepackage{array}
\usepackage{dsfont}

\newcommand{\ignore}[1]{}  

\newcommand{\hermitian}[0]{\text{H}}
\newcommand{\transpose}[0]{\text{T}}

\newcommand{\Ts}[0]{T_{\text{s}}}
\newcommand{\eF}[0]{\epsilon_{\text{F}}}
\newcommand{\eT}[0]{\epsilon_{\text{T}}}
\newcommand{\Nt}[0]{N_{\text{T}}}
\newcommand{\Nr}[0]{N_{\text{R}}}
\newcommand{\Nzc}[0]{N_{\text{zc}}}

\begin{document}
\title{
Software Defined Radio Implementation of Carrier and Timing Synchronization for Distributed Arrays}

\author{%
Han Yan\\ 
Dept. of Electrical and Computer Eng.\\
University of California, Los Angeles\\
yhaddint@ucla.edu
\and
Samer Hanna\\ 
Dept. of Electrical and Computer Eng.\\
University of California, Los Angeles\\
samerhanna@ucla.edu
\and
Kevin Balke\\ 
Google Inc.\\
fughilli@gmail.com
\and 
Riten Gupta\\
UtopiaCompression Corporation\\
riten@utopiacompression.com
\and
Danijela Cabric\\
Dept. of Electrical and Computer Eng.\\
University of California, Los Angeles\\
danijela@ee.ucla.edu
\thanks{\footnotesize 978-1-5386-6854-2/19/$\$31.00$ \copyright2019 IEEE}              
}

\maketitle

\thispagestyle{plain}
\pagestyle{plain}

\maketitle

\thispagestyle{plain}
\pagestyle{plain}

\begin{abstract}
The communication range of wireless networks can be greatly improved by using distributed beamforming  from a set of independent radio nodes. One of the key challenges in establishing a beamformed communication link from separate radios is achieving carrier frequency  and sample timing synchronization. This paper describes an implementation that addresses both carrier frequency and sample timing synchronization simultaneously using RF signaling between designated master and slave nodes. By using  a pilot signal transmitted by the master node, each slave estimates and tracks the frequency and timing offset and digitally compensates for them. A real-time implementation of the proposed system was developed in GNU Radio and tested with Ettus USRP N210 software defined radios. The measurements show that the distributed array can reach a residual frequency error of 5 Hz and a residual timing offset of 1/16 the sample duration for 70 percent of the time. This performance enables distributed beamforming for range extension applications.
\end{abstract}

\tableofcontents

\section{Introduction}

Cooperation among geographically distributed radios, aka distributed arrays, provides a performance boost in a variety of applications. Distributed beamforming (DBF) is a nascent wireless communications technology that seeks to offer signal power gains using distributed arrays. With distributed transmitter and receiver beamforming, groups of radios transmit messages over a longer distances and more power-efficiently than any single node pair could achieve \cite{Brown_2013_DBF_DEMO,Hayes_2016_DBF_DEMO}. With $\Nt$ nodes in a transmit group and $\Nr$ receive nodes, a joint transmit/receive beamforming system can theoretically achieve a potential gain improvement of $\Nt^2\Nr$ relative to a point-to-point link. The extremely concentrated energy field from DBF can also be used for far-field wireless energy delivery \cite{reynolds_2016_EH_DEMO} which reduces energy in unintended areas where humans could be adversely affected by microwaves \cite{RZhang_2017_EHSurvey}. Distributed multiple-input multiple-output (DMIMO) is another application of distributed arrays which relies on the increased virtual array dimension to provide the required degrees-of-freedom for spatial multiplexing and interference control. This technique facilitates significant spectral efficiency improvement \cite{Caire_2015_DMIMO_DEMO}, \cite{Irmer_2011_CoMP}. A distributed array can also synthesize a larger virtual aperture and provide orders-of-magnitude higher precision in localization \cite{Win_2009_DLocalization} and radar imaging \cite{Reynolds_2017_DImaging} applications.

Synchronization among radios is the key requirement for distributed beamforming, MIMO, localization and imaging, since carrier frequency offset/drift, and sample timing offset introduce non-coherence and severely degrades performance. 
Synchronization schemes in most recent prototype studies \cite{Hayes_2016_DBF_DEMO,reynolds_2016_EH_DEMO,Caire_2015_DMIMO_DEMO,Reynolds_2017_DImaging} mostly rely on dedicated cables, e.g., copper and optical fiber, or GPS clocks. The over-the-wire reference signals, e.g., 10 MHz and pulse-per-second (PPS) signals, greatly simplify the synchronization, but they are not suitable for mobile systems. Meanwhile, reliance on GPS clocks may greatly increase the cost of the radios. A digital signal processing (DSP) based in-band synchronization protocol is an alternative solution, favorable due to its potentially low complexity and cost in synchronizing distributed arrays. DSP oriented synchronization is especially suitable for distributed mobile radios, e.g., cooperative unnamed aerial vehicle access \cite{UAV_CoMP} and backhaul \cite{UAV_cooperative_backhaul}. In this work, we focus on the in-band carrier frequency and sample timing synchronization.

Synchronization among distributed radio nodes has been analyzed theoretically; for example, see the survey papers \cite{DMIMO_survey,dist_sync_survey} and references therein. Recently, a number of software-defined-radio (SDR) proof-of-concept implementations have been reported. Works \cite{UCSB_2013_TWC,UCSB_2016_TWC} present a novel system for transmitter or receiver beamforming where slave radios estimate a continuous wave (CW) signal 
from the master node and use this to estimate the carrier frequency offset and adjust the phases of baseband samples to reach synchronization. 
In \cite{Brown_2017_ICASSP,Brown_2018_aeroconf} an SDR implementation of sample timing is
developed while using perfectly synchronized carriers with 10 MHz references. Reference  \cite{Qiu_2016_MobiHoc} considers a broadcast system and sub-$\mu$s level timing alignment is achieved for the WiFi OFDM waveform. 
An in-band synchronization protocol is developed in \cite{katabi_2012_DMIMO,Caire_2013_AirSync}
which results in a performance boost, as measured by throughput. While these studies have
advanced the state-of-the-art in DBF, there is a scarcity of studies on SDR implementation of {\it joint frequency and timing synchronization} that supports real-time DBF and DMIMO data communications.

In this work, we focus on the real-time SDR implementation of joint carrier and timing synchronization for DBF applications using the Universal Software Radio Peripheral (USRP)\cite{usrp_n210}. We develop a lightweight protocol for distributed radio nodes to achieve synchronization with a master-slave architecture. As opposed to \cite{katabi_2012_DMIMO,Caire_2013_AirSync}, we focus on synchronization accuracy rather than throughput improvement. We use the theoretical performance bound of synchronization accuracy as well as required accuracy for DBF from our simulation as guidelines in protocol design. The experimental results show that the proposed scheme achieves under 5 Hz residual frequency error. The timing accuracy can be lower than $1/16$ of the sample duration nearly 70-percent of the time, based on an over-night trial. Our proposed protocol and algorithm are tailored for DBF applications, but are straightforward to extend to other applications as well. 

The rest of the paper is organized as follows. In Section~\ref{sec:system_model}, we introduce the concept of distributed arrays and the system model for synchronization. 
We present the proposed synchronization protocol and DSP algorithms in Section~\ref{sec:protocol} and \ref{sec:algorithm}, respectively. The theoretically achievable and the required synchronization accuracy analysis for practical scenarios is presented in Section~\ref{sec:accuracy_analysis}. SDR implementation details are discussed in Section~\ref{sec:implementation} and experimental results, which verify our analysis, in Section~\ref{sec:results}. Section~\ref{sec:discussion} discusses the limitations of the current implementation and future research directions. Finally, Section~\ref{sec:conclusion} concludes the paper.

\textit{Notations:} Scalars, vectors, and matrices are denoted by non-bold, bold lower-case, and bold upper-case letters, respectively, e.g. $h$, $\mathbf{h}$ and $\mathbf{H}$. The element in the $i$-th row and $j$-th column in matrix $\mathbf{H}$ is denoted by $\{\mathbf{H}\}_{\mathit{}i,j}$. Conjugate, transpose and Hermitian transpose are denoted by $(\cdot)^*$, $(\cdot)^{\transpose}$ and $(\cdot)^{\hermitian}$, respectively. 

%
%
\section{System Model}
\label{sec:system_model}
In this section, we introduce the concept of distributed array systems and we present the challenges in achieving synchronization between radios. The nomenclature used in the paper is summarized in Appendix~\ref{app:notation}.

\subsection{Distributed array system concept}
\begin{figure}
\centering
\includegraphics[width=3.4in]{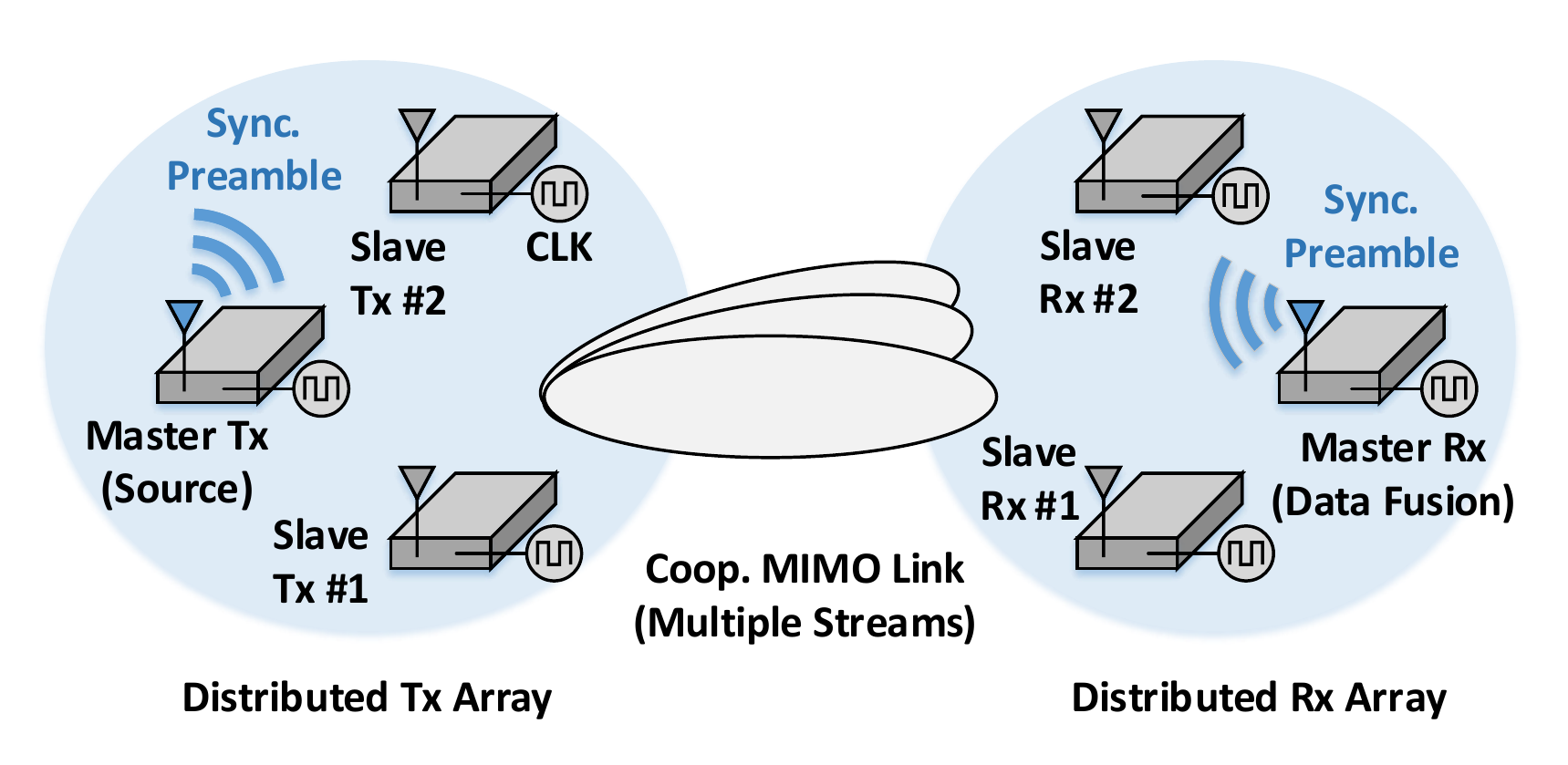}
\caption{\bf{Cooperative MIMO communication with an example of $\Nt = 3$ and $\Nr = 3$ distributed array.}}
\label{fig:DMIMO_system}
\end{figure}

Consider two groups consisting of single antenna radios, as shown in Figure~\ref{fig:DMIMO_system}, trying to communicate wirelessly.
Each radio has its own clock for carrier synthesis and sample timing. In order for each group to behave as an antenna array, thus performing cooperative communication, all the radios within the group need to be synchronized. 

The system we propose uses the same frequency band for synchronization and cooperative communication, and these two functions are interleaved in time. We use a master-slave based intra-group synchronization architecture which is suited to reduce the synchronization time for a  small scale group having direct links between the master and the slaves. In the synchronization phase, the master broadcasts a preamble and the slaves use it to estimate the frequency and timing offsets between themselves and the master. The slaves then compensate for the offsets using baseband processing in order to enable cooperative communication.

\subsection{System model for synchronization}


Let us consider the synchronization between the master and one of the  slaves. The results obtained can easily be  extended to multiple slaves due to the master-slave architecture. We focus on a narrow-band system and leave the wide-band system for future work. 

Due to the use of multiple oscillators and timing signals, there is a sample timing offset\footnote{In this work, $\Delta \tau$ includes the contribution of both the hardware timing offset and propagation delay.} (STO) $\Delta \tau$, with units of seconds, and a carrier frequency offset (CFO) $\Delta f$, with units of Hz, between the master and each slave. The sample duration $\Ts$ is identical\footnote{In this work, we do not model and compensate the different sample clock skew among radios. It is a valid assumption as explains in footnote 5.} in the master and the slave. For notational convenience, we define the normalized CFO as $\eF=2\pi \Delta f\Ts$.

Let us denote the transmitted signal from the master as $s[n]$. Then the digitized signal at the slave can be expressed as 
\begin{align}
\begin{split}
r[n] 
=& h_0e^{j\eF n}\sum_{k=-\infty}^{+\infty}s[k] p_{\text{ps}}\left((n-k)\Ts-\Delta \tau\right)+w[n].
\end{split}
\label{eq:slave_rx_signal1}
\end{align}
where the complex channel gain and hardware phase offset between master and slave is denoted as $h_0$, the thermal noise at the slave radio is modeled as additive white Gaussian noise (AWGN) $w[n]\sim\mathcal{CN}\mathrm{}(0,\sigma_{\text{n}}^2)$, and the signal to noise ratio is defined as $\text{SNR}_{\text{sync}} = |h_0|^2/\sigma_{\text{n}}^2$. Due to the delay $\Delta \tau$, there is a misalignment in transmitter and receiver sample ticks. Therefore, continuous-time analog filtering\footnote{Analog filtering includes the baseband filter after the digital-to-analog converter and the passband filter after the power amplifier.}
 at the transmitter is used and is modeled as the function $p_{\text{ps}}(t)$. The exact expression of such a filter is hardware dependent and is commonly unknown in SDR implementations. For tractable algorithm design in this work, we utilize the approximation that $p_{\text{pc}}(t)$ is a known low-pass interpolation function. 

The objective of synchronization is to design a protocol and a signal processing algorithm which
together allow the slave to use the received signal  to
estimate the CFO $\Delta f$ and STO $\Delta \tau$.


%
%
\section{Proposed protocol}
\label{sec:protocol}
 In this section, we introduce the proposed frame structure as well as the procedure for synchronization. In order to quantify the required overhead in synchronization, the cooperative communication procedure is also briefly introduced.


\subsection{Frame structure}
\begin{figure}
\centering
\includegraphics[width=3.4in]{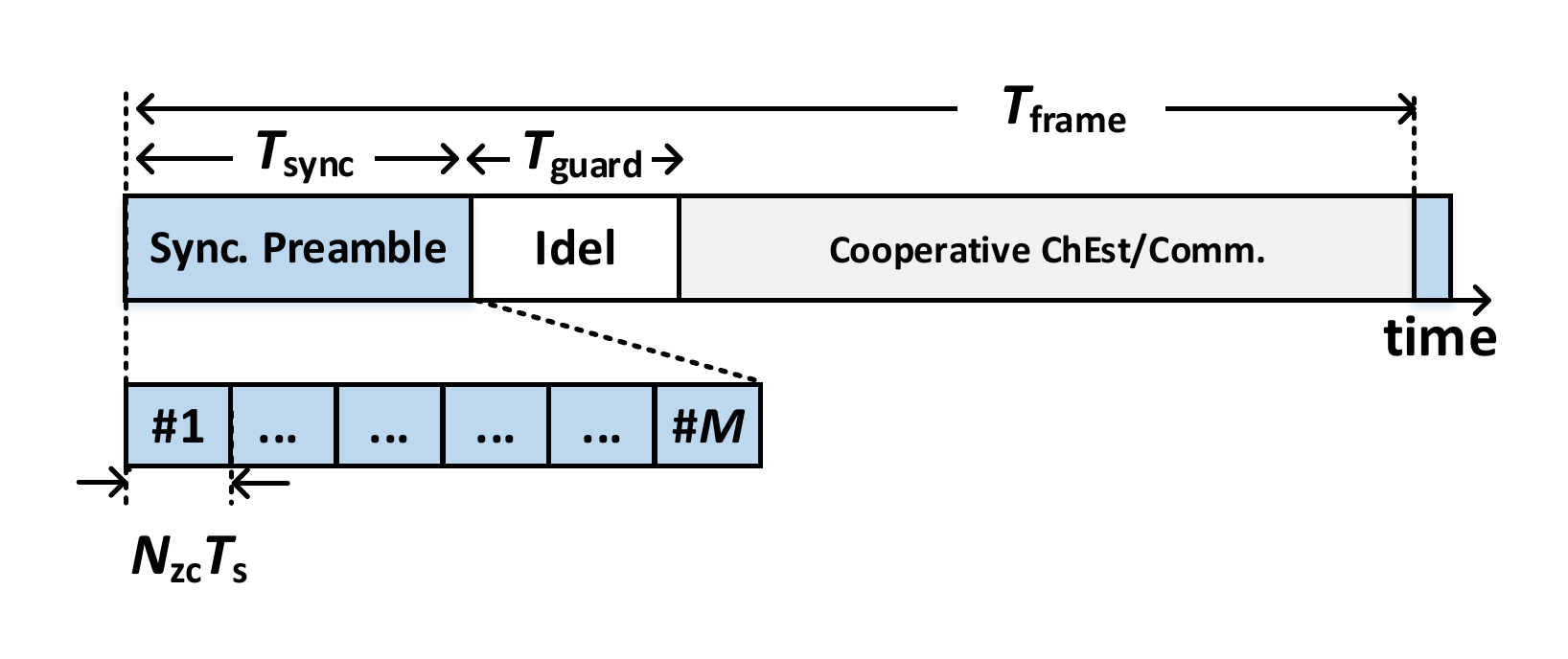}
\caption{\bf{Proposed frame structure for synchronization and cooperative communication.}}
\label{fig:frame_structure}
\end{figure}

The proposed frame structure is shown in Figure~\ref{fig:frame_structure}. The frame contains a periodic synchronization preamble and cooperative communication segment with period $T_{\text{frame}}$. This value depends on the drift rates of frequency and timing offsets, which requires calibration of the hardware.

Due to constraints of implementation as well as overhead efficiency concerns, the proposed frame does not include a transmitter time stamp as is commonly utilized in distributed embedded systems. We propose to use an $M$-repetition ($M\geq 2$) Zadoff-Chu (ZC) sequence \cite{ZC_sequence}, $s_{\text{zc}}[n],n\in[0,\Nzc-1]$ with length $N_{\text{zc}}$, as the synchronization preamble. The ZC sequence is known for its perfect cyclic-autocorrelation property, i.e., 
\begin{align}
\sum_{n=0}^{\Nzc-1}s_{\text{zc}}[n]s^{*}_{\text{zc}}[\text{mod}(n+d,\Nzc)] =
\begin{cases}
\Nzc,&d=0\\
0,& 0<d< \Nzc
\end{cases}.
\end{align}
Note that the duration of the synchronization preamble is $T_{\text{sync}} = MN_{\text{zc}}T_{\text{s}}$, where $T_{\text{s}}$ is the sample duration. 
The synchronization preamble used in (\ref{eq:slave_rx_signal1}) is expressed as
\begin{align}
s[n] = \sum_{m=0}^{M-1}s_{\text{zc}}[n-mN_{\text{zc}}].
\label{eq:s_of_n_inf_sequence}
\end{align}

A fixed time duration $T_{\text{guard}}$, known to both slave and master, is reserved for the slave nodes for baseband processing. The specific value depends on hardware implementation, e.g., the propagation delay, radio-frequency (RF) and baseband (BB) processing delay.

The cooperative communication interval is divided into three phases, intra-group channel estimation (ChEst), inter-group ChEst, and data communication. The duration of inter-group ChEst $T_{\text{inter}}$ depends on the channel variation due to environment changes. The length $T_{\text{intra}}$ of intra-group ChEst depends on the intra-group channel variation due to residual synchronization error and is used in evaluating overhead in Section~\ref{sec:results}.




\subsection{Procedure}
The master radio obeys the frame structure defined above, and implements it using its own clock. 

In an asynchronous manner within the group, each slave radio actively searches for the synchronization preamble. Upon detection of the preamble, the slave uses the received signal to estimate the carrier and timing offsets relative to the master, and use their estimator $\Delta \hat{f}$ and $\Delta \hat{\tau}$ to adjust baseband signal during the cooperative communication period that follows. Specifically, the slave utilizes the known idle period for necessary processing to adjust the baseband signal such that the transmitted and received inter-group signals are synchronized. The slave radio also utilizes the periodic nature of the preamble to retain tight synchronization as well as improve synchronization accuracy by filtering over the estimated offset across multiple repetitions of the ZC sequence.  

Cooperative communication occurs when the synchronization is reached within a specified accuracy range, i.e., residual synchronization error.  As such, the cooperative communication stage has residual CFO $\Delta_{\text{R,F}} \triangleq \Delta f- \Delta \hat{f}$ and STO $\Delta_{\text{R,T}} \triangleq \Delta \tau- \Delta \hat{\tau}$, which is much smaller than the original errors $\eF$ and $\eT$.



\section{Proposed Synchronization Algorithms}
\label{sec:algorithm}
In this section, we present the proposed DSP based synchronization algorithm used by the slave radios.

For notational convenience 
we first re-write the STO as $\Delta \tau = \eT \Ts=(d^{\star}+\zeta^{\star})\Ts$, where $\eT$ is the sample-duration-normalized timing offset, with integer part $d^{\star} = \lfloor \Delta\tau/\Ts \rfloor$ and fractional part $\zeta^{\star} = \eT - d^{\star}$, i.e., $\zeta^{\star}\in [0,1)$.  The $\Ts$-normalized residual timing error is denoted as $\epsilon_{\text{R,T}} \triangleq \Delta_{\text{R,T}}/\Ts$. Also, we define the normalized CFO as $\eF=2\pi \Delta f\Ts$.

The received signal (\ref{eq:slave_rx_signal1}), after plugging in the ZC sequence expression and using the above notational modification, is rewritten as
\begin{align}
\begin{split}
r[n] 
=& h_0e^{j\eF n}\bigg[\sum_{k=-\infty}^{+\infty}\sum_{m=0}^{M-1}s_{\text{zc}}[k-mN_{\text{zc}}]\\
&\cdot p_{\text{ps}}\left(\left(n-d^{\star}-k\right)\Ts-\zeta^{\star}\Ts\right)\bigg]+w[n].
\end{split}
\label{eq:slave_rx_signal_raw}
\end{align}

\subsection{Preamble Detection and Integer Sample Timing Recovery}
The zero-autocorrelation property of the ZC sequence provides a simple method for preamble detection. Define the null and alternate hypotheses $\mathcal{H}_{\mathrm{0}}:$ preamble absent and $\mathcal{H}_{\mathrm{1}}:$ preamble present. We form a test statistic sequence $y_{\text{corr}}[n]$ by
correlating the ZC sequence $s_{\text{ZC}}$ with $M$ length-$\Nzc$ segments of the received signal $r[n]$ and combining the results as shown below.
%
\begin{align}
y_{\text{corr}}[n] \triangleq \sum_{m=0}^{M-1}\left|\sum_{k=0}^{N_{\text{zc}}-1}s_{\text{zc}}[k]r^*[n+k+mN_{\text{zc}}]\right|^2 \underset{\mathcal{H}_{\mathrm{0}}}{\overset{\mathcal{H}_{\mathrm{1}}}{\gtrless}} \eta
\end{align}
where $\eta$ is the detection threshold whose optimal value can be calculated based on the thermal noise power. The 2-norm summation over $M$ adjacent $N_{\text{zc}}$ windows is to avoid phase rotation from frequency offset\footnote{The normalized CFO is up to $\eF \approx 0.075 \text{ rad}$ based on the maximum USRP frequency offset of $\pm5$ppm, when using a 2.4GHz carrier frequency, and a 1MHz sampling rate. The complex envelop due to CFO has its phase rotating with values larger than $2\pi$, which severely degrades the correlation peak if we used a long ZC sequence as compared to multiple repetitions of a shorter sequence. With such a setting, the $\pm$5 ppm sample clock skew introduces $0.05\Ts$ STO drift within 10ms and it is negligible in a DBF application as shown later in our simulations.}. 

Moreover, based on the location of the correlation peak, the integer timing offset $d^{\star}$ can be estimated via
\begin{align}
\hat{d}^{\star} = \text{arg} \max_n y_{\text{corr}}[n]
\label{eq:integer_delay_estimator}
\end{align}

It is worth noting that the perfect autocorrelation property of ZC sequences is degraded due to fractional-sample delay between master and slave. In other words, when $\zeta^{\star}\approx 0.5$ a small variation in $y_{\text{corr}}[n]$ due to AWGN causes the correlation peak to lock either to $\hat{d}^{\star} = d^{\star}$ or $\hat{d}^{\star} = d^{\star}+1$.
To deal with this, a fractional delay estimator must be designed and implemented, as we describe later.

\subsection{CFO Estimation and Compensation}
We start with a one shot estimation algorithm which utilizes the known repetition pattern of the preamble signal for CFO estimation. Due to the fact that the received signal within the sample window $[\hat{d}^{\star},\hat{d}^{\star}+M\Nzc -1]$, i.e., the estimated time window for $M$ consecutive ZC sequences in terms of the slave's clock, is periodic with period $\Nzc$, we propose to use 
\begin{align}
\Delta \hat{\epsilon}_{\text{F}} = \frac{\angle \left(\sum_{n=0}^{(M-1)N_{\text{zc}}-1}r[\hat{d}^{\star}+N_{\text{zc}}+n]r^*[\hat{d}^{\star}+n]\right)}{N_{\text{zc}}}
\label{eq:CFO_est}
\end{align}
as the normalized CFO estimate.  

The estimate for $\Delta f$ is available by straightforward scaling
of $\Delta \hat{\epsilon}_{\text{F}}$. The angle operator $\angle(\cdot)$ returns the phase of a complex number and can be computed by $\angle(x) = \text{tan}^{-1}[\Im(x)/\Re(x)]$. Furthermore, this approach intrinsically assumes that the phase rotation due to CFO is unwrapped, i.e., $\eF \Nzc<2\pi$. With the value $N_{\text{zc}}=63$ this condition is easily met (see footnote 5)
. For applications in which a longer ZC sequence is required for more processing gain, \cite{madhow_phase_wrapped} provides an algorithm for wrapped phase measurement.

Due to the fact that our system uses a dedicated preamble for CFO estimation rather than an unmodulated tone \cite{UCSB_2013_TWC}, the delay error is crucial in the design. The reason for using auto-correlation in (\ref{eq:CFO_est}) is its resilience to the unknown fractional delay, i.e., in the high $\text{SNR}_{\text{sync}}$ regime, $r[n+\hat{d}^{\star}+N_{\text{zc}}]r^{*}[n],\forall n\in[0,(M-1)N_{\text{zc}}-1]$ gives an estimate of $e^{j\eF N_{\text{zc}}}$ regardless of $\eT$ and the potential 1-sample ambiguity in $\hat{d}^{\star}$. In contrast, approaches which rely on directly decoding $s_{\text{zc}}$ cannot avoid estimating the fractional delay for CFO estimation. 

The CFO estimate can be filtered to provide more accuracy. The extended Kalman filter (EKF) is commonly used as an averaging filter \cite{UCSB_2013_TWC}. The state vector consists of the phase offset $\phi$ and normalized CFO $\eF$ between the master and slave. Assuming the channel variation $h_0$ is compensated, the state update function is expressed as the following linear equation
\begin{align}
\underbrace{\begin{bmatrix}
\phi_{k}\\
\epsilon_{\text{F},k}
\end{bmatrix}}_{\mathbf{x}_{\mathnormal{k}}} = 
\underbrace{\begin{bmatrix}
1 &N_{\text{CYC},k} \\
0 & 1
\end{bmatrix}}_{\mathbf{F}} 
\underbrace{\begin{bmatrix}
\phi_{k-1}\\
\epsilon_{\text{F},k-1}
\end{bmatrix}}_{\mathbf{x}_{\mathnormal{k-1}}} +
\underbrace{\begin{bmatrix}
\delta \phi_k\\
\delta \epsilon_{\text{F},k}
\end{bmatrix}}_{\mathbf{v}_{\mathnormal{k}}}
\end{align}
where we add subscript $k$ to indicate the $k$-th preamble for clarity and we use this notation in the remainder of this subsection. Note that $N_{\text{CYC},k}$ is the number of samples between the $k$-th and the $(k+1)$-th preambles. By default $N_{\text{CYC},k} = T_{\text{CYC}}/\Ts,\forall k$ if the master hardware timer is perfect. In practice, slaves should retain a timer to measure the actual time gap between adjacent synchronization preambles. The vector $\mathbf{v}_{\mathnormal{k}}$ contains the phase drift and frequency drift values at time instant $k$.

The observation vector contains the indirect measurement of the phase offset estimator and CFO estimator from (\ref{eq:CFO_est}). 
\begin{align}
\underbrace{\begin{bmatrix}
\cos(\hat{\phi_k})\\
\sin(\hat{\phi_k})\\
\hat{\epsilon}_{\text{F},k}
\end{bmatrix}}_{\mathbf{z}_{\mathnormal{k}}} = 
\underbrace{\begin{bmatrix}
\cos(\phi_k)\\
\sin(\phi_k)\\
\epsilon_{\text{F},k}
\end{bmatrix}}_{g(\mathbf{x}_{\mathnormal{k}})} +
{\mathbf{w}_{\mathnormal{k}}}
\end{align}
Specifically, we use $\cos(\hat{\phi_k}) = \Re(r[\hat{d}_k^{\star}])$ and $\sin(\hat{\phi_k}) = \Im(r[\hat{d}_k^{\star}])$ from the actual received signal. The vector $\mathbf{w}_{\mathit{k}}$ contains measurement errors whose variances are determined in an empirical manner. In processing the preamble, the slave filters the estimated CFO via standard Kalman Filter steps \cite{sayed2011adaptive} and provides a CFO estimator $\hat{\epsilon}_{\text{F}}$ with better precision.

The CFO in the received signal can be compensated via
\begin{align}
\tilde{r}[n] = r[n]e^{-j\hat{\epsilon}_{\text{F}}n}.
\end{align}
and the following baseband signal is used in cooperative communication. It is worth noting that the above approach is not intended to adjust the phase offset.

\subsection{Fractional Sample Timing Recovery}
Since the correlation and peak detector only provide coarse timing information, i.e., nearest receiver sample instance, the accuracy is largely compromised. We propose to use a maximum likelihood estimator for fractional delay estimation and utilize the fact that the search range is within $[0,\Ts)$ of the estimated integer delay. 

Due to the fact that the integer delay estimator $\hat{d}^{\star}$ is sensitive to AWGN when $\zeta^{\star}$ is close to 0.5, an approach for dealing with such errors is required. We propose to use a simple approach and set the fractional delay candidate to be within $[-0.5\Ts,0.5\Ts)$ rather than $[0,\Ts)$ in estimation. This facilitates automatically compensating integer estimation error. 

For this purpose, the digitized signal is correlated with a
bank of filters, each of which is a fractionally delayed ZC sequence. 
\begin{align}
\hat{\zeta}^{\star} = \text{arg} \max_{\zeta\in\mathcal{D}} \sum_{m=0}^{M-1}\left|\sum_{n=0}^{N_\text{zc}-1}\tilde{r}^*[n+\hat{d}^{\star}+mN_{\text{zc}}]\tilde{s}^{(\zeta)}_{\text{zc}}[n]\right|^2
\label{eq:fractional_delay_estimator}
\end{align}
where the $\zeta$-fractionally delayed ZC sequence $r^{(\zeta)}_{\text{zc}}[n]$ is obtained from linear intepolation, i.e.,
\begin{align}
\tilde{s}^{(\zeta)}_{\text{zc}}[n] = \sum_{k=0}^{\Nzc - 1} s_{\text{zc}}[\text{mod}(n-k,\Nzc)]p_{\text{pc}}(t-\zeta-k\Ts)
\end{align}
Note that the optimal design requires knowledge of the transmitter RF filter, $p_{\text{ps}}(t)$, but mismatch in this knowledge causes only minor degradation. Practically, $N_{\zeta}$ fractionally-delayed ZC sequences are used and $\zeta$-candidates are chosen from dictionary with a step-size $\Ts/(N_{\zeta}+1)$, i.e., 
\begin{align}
\mathcal{Z} = \mathrm{}\left\{\frac{-\mathit{}\Ts}{2},\mathrm{}\frac{-\mathit{}\Ts}{2}+\mathit{}\frac{\Ts}{N_{\zeta}+\mathrm{}1}, \cdots, \frac{\mathit{}-\Ts}{\mathrm{}2}+\frac{\mathit{}N_{\zeta}\mathit{}\Ts}{N_{\zeta}+\mathrm{}1}\right\}
\end{align}
and we choose $N_{\zeta}$ according to the residual STO target.


\subsection{Computational complexity}

The complexity is divided into two parts: always-on preamble detection and detection triggered synchronization algorithm. The detection algorithm uses an $N_{\text{zc}}$-tap finite impulse response (FIR) filter and therefore requires $MN_{\text{zc}}$ complex multiplications and accumulations at each sample. Once the preamble detection is triggered, a total of $MN_{\text{zc}}$ samples are used for additional processing, i.e., CFO estimation and fractional delay estimation. The former requires $(M-1)N_{\text{zc}}$ complex multiplications and accumulations, 1 division and phase computation. The latter requires $N_{\zeta}$ pre-computed $\tau$-delayed ZC sequences. Correlation with each sequence requires $MN_{\text{zc}}$ complex multiplications and accumulations and requires a total of $MN_{\text{zc}}N_{\zeta} $ complex operations.

\section{Achievable and Target Accuracy Analysis}
\label{sec:accuracy_analysis}

In this section, we start with a discussion of the theoretically achievable synchronization offset estimator variance. Then, we use simulations to study the impact of residual synchronization error on DBF performance. This analysis serves as a guideline in evaluating our algorithm and SDR implementation.

\subsection{Analysis of achievable accuracy}
Reference \cite{Brown_2012_timing_CRLB} shows that the maximum likelihood estimator achieves the Cram\'er-Rao lower bound but such an estimator actually estimates the accumulation of hardware delay and propagation delay and does not distinguish between them.

From the Cram\'er-Rao lower bound perspective, the variance of CFO estimation and STO estimation are 
\begin{align}
\mathrm{var}(\Delta\mathnormal{\hat{f}}) \geq \text{CRLB}(\mathrm{}\Delta \mathit{}\hat{f}) = \mathrm{}\frac{3}{2\pi^2\mathit{T}\mathrm{}^2_{\text{est}}\text{SNR}_{\text{sync}}}
\label{eq:CFO_CRLB}
\end{align}
\begin{align}
\mathrm{var}(\Delta\mathnormal{\hat{\tau}}) \geq \text{CRLB}(\mathrm{}\Delta \mathit{}\hat{\tau}) = \mathrm{}\frac{12\pi \mathit{T}_{\text{s}}\mathrm{}^3}{\mathit{T}_{\text{est}}\text{SNR}_{\text{sync}}}
\label{eq:STO_CRLB}
\end{align}

where $T_{\text{est}} = M\Nzc \Ts$ is the observation duration \cite{freq_CRLB},\cite{timing_CRLB}. Consider a setting with $\Ts=1$ $\mu\text{s}$ and $M=10$ repetitions of ZC sequence, i.e., $T_{\text{est}} = 0.63$ ms, and 20 dB SNR. The theoretical CFO estimator standard deviation is  $61.9$ Hz. With the help of a Kalman filter the accuracy, with averaging over multiple preambles, can be improved. On the other hand, the theoretical fractional delay estimation standard deviation is $0.008\Ts$, a specification good enough for cooperative communication as shown in section~\ref{sec:results}. This implies that filtering of the delay estimator over multiple preambles is not necessary.

It is worth noting that the proposed approach for timing estimation does not incorporate the propagation delay between master and slaves. In other words, the estimator is not unbiased and in the proposed approach $\mathbb{E}|\mathit{\epsilon}_{\text{R,T}}| \geq L/c$ where $L$ is the longest intra-group distance between master and slaves. This design is intended for small scale distributed cooperative communication and therefore the propagation delay within the group is not addressed. The work \cite{Brown_2018_aeroconf} provides an opposite way of timing estimator design where closed-loop forward and backward links are used to address the intra-group propagation delay at the cost of having to synchronize each master/slave pair independently, rather than joint synchronization by broadcasting. 

\subsection{Simulation of synchronization requirements}
First we use simulations to study the impact of residual synchronization error on system performance. We consider a distributed beamforming system for range extension in line of sight (LOS) channels between $\Nt$ transmit and $\Nr$ receive radios. The inter-group distance is $20$ km. The intra-group radio distance is 50 m on average, and the LOS environment features a very sparse channel with rank 1. The SNR between any pair of transmitter and receiver radios is -1.5 dB\footnote{Consider 23dBm transmission power from each transmitter radio, free space pathloss model with 2.0 GHz carrier frequency, 10 MHz noise bandwidth, and 4dB noise figure.}. Simulations were conducted with perfect phase adjustment at the beginning of the cooperative communication sub-frame, i.e., $\Nt^2\Nr$ beamforming gain. The gain drops as the phase coherence degrades due to residual frequency offset. We use the Signal to Interference and Noise Ratio (SINR) as a metric to evaluate the coherence of the system. The interference is due to the intersymbol interference resulting from the timing synchronization errors among symbols. 

We evaluate the performance of the system under Gaussian distributed frequency or timing residue errors individually. From this evaluation, we estimate the required accuracy of synchronization. To achieve a Bit Error Rate (BER) of $10^{-5}$ using 16 QAM modulation, we need an SINR of at least 20 dB. We wish to determine the highest acceptable RMS residual timing and frequency errors to achieve this SINR. 

\begin{figure}
\centering
\includegraphics[width=3.4in]{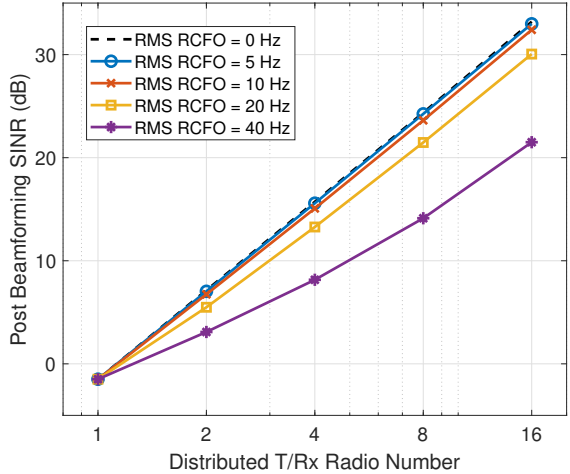}
\caption{\bf{Impact of Gaussian Distributed residual carrier frequency error on distributed beamforming gain}}
\label{fig:required_RCFO}
\end{figure}

\begin{figure}
\centering
\includegraphics[width=3.4in]{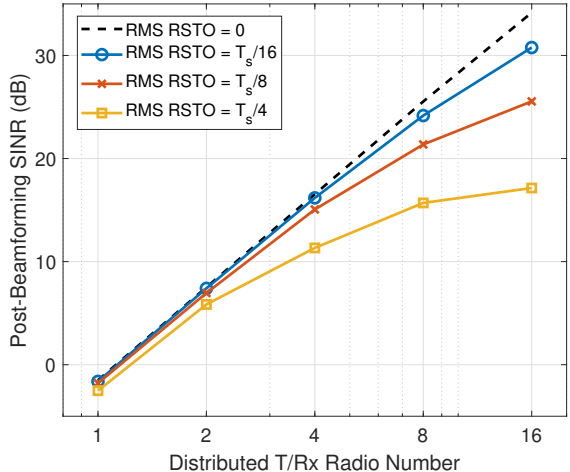}
\caption{\bf{Impact of Gaussian Distributed residual timing error on distributed beamforming gain.}}
\label{fig:required_RSTO}
\end{figure}

For frequency errors, Figure~\ref{fig:required_RCFO} shows the post beamforming SINR at the end of a frame where cooperative communication duration is $5$ ms. In this figure, the x-axis refers to the number of transmitter ($\Nt$) and receiver ($\Nr$) radios. The results indicate that to achieve a  20 dB SINR using a  group consisting of $\Nt=\Nr=8$ radios, we need the  frequency RMS error $\Delta_{\text{R,F}} = \Delta f-\Delta \hat{f}$ to be within 20 Hz.  As for the timing with the same group size, Figure~\ref{fig:required_RSTO} shows the post beamforming SINR  is greater than 20 dB when the RMS residual timing error is within $\Ts/8$.

%
%
\section{Software-Defined Radio Implementation}
\label{sec:implementation}
The proposed protocol and the synchronization algorithm were implemented using the USRP N200/N210 software defined radio kits. In our system, we used 5 USRPs. One USRP was used as a master, two USRPs as slaves, and two USRPs acted as receivers whose purpose was to evaluate the performance of the system.  In this section, we start by introducing the capabilities of the SDR kits. Then we describe the parameters and details of our implementation.
\begin{figure*}[th]
\centering
\includegraphics[width=6.5in]{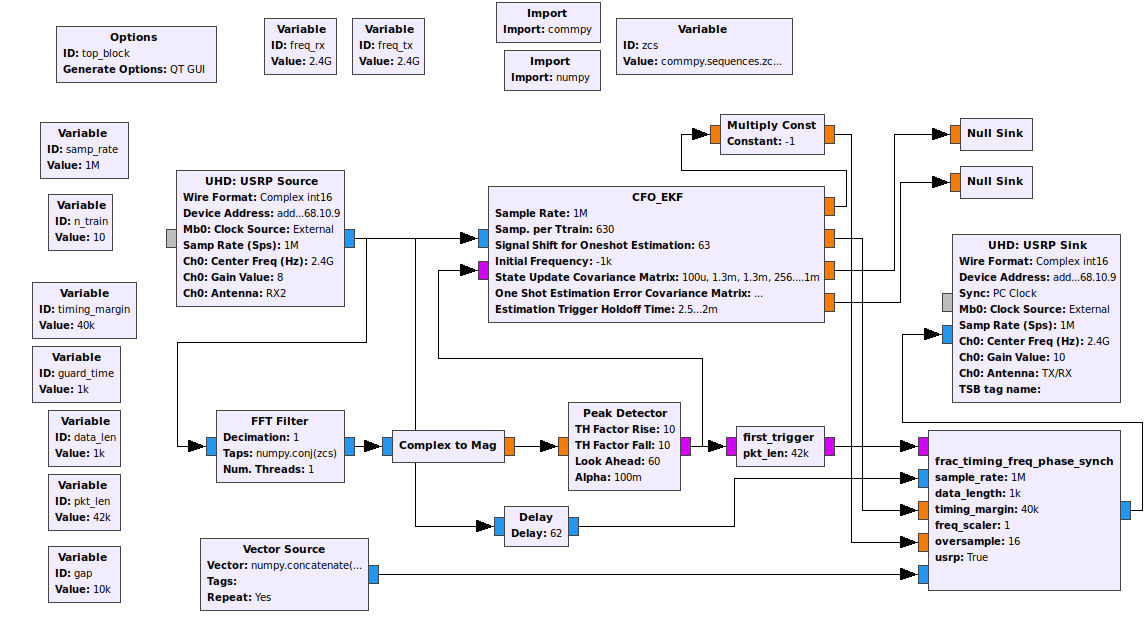}
\caption{\bf{Algorithm flow-chart in our USRP implementations with Gnuradio.}}
\label{fig:USRP_setup}
\end{figure*}
\subsection{Introduction to the USRP N200/N210 platform}
The Ettus USRP N200/N210 series is a software defined radio kit designed for RF applications from DC to 6 GHz.  The RF capabilities of the USRP are determined by the installed RF daughter-board. Our implementation used a set of heterogeneous RF daughter-boards including the SBX having an RF frontend that can operate in the frequency range from 0.4 to 4.4 GHz, and the XCVR2450 which is operational from 2.4 to 2.5  and 4.9 to 6.0 GHz. 
 Besides the RF frontend, the USRP consists of an  analog-to-digital converter (ADC) and digital-to analog converter (DAC) as well as a field-programmable gate array (FPGA).
 The FPGA implements some simple digital signal processing such as upsampling and downsampling signals to the rates required by the ADC/DAC. It is also used to communicate with the host computer. On the host, the USRP Hardware Driver (UHD) provided by Ettus is used to interface the host to the USRP.  Digital signal processing software such as  GNU Radio can be used to operate the USRP. 
 Each USRP has its own oscillator and its own timing clock. In order to support MIMO communications, three options are available for carrier and timing synchronization \cite{usrp_mimo}. These options include (1) a MIMO cable, which lets two USRPs share both frequency and timing signals, (2) using an external 10 MHz carrier reference clock and pulse-per-second (PPS) timing synchronization signal, or (3) using a GPS Disciplined Oscillator  (GPSDO) \footnote{According to its datasheet \cite{USRP_GPSDO}, the GPSDO module  has $\pm50$ nanosecond accuracy in the clock timing and $\pm0.025$ ppm frequency accuracy, which translates to $\pm60$ Hz carrier frequency offset when operating at 2.4 GHz band.}. Except for the GPSDO, the other two options require external connections which are not suitable for a distributed system. As for the GPSDO, its accuracy depends on the ability of the USRP to receive a GPS signal, which renders it unsuitable for indoor deployments, and gives limited frequency synchronization ability.

It is worth noting that the USRP internal clock is not stable as it retunes the frequency after each transmission and this results in abrupt frequency changes.
This phenomenon is a result of the carrier generator design and could have been avoided by alternate design choices.  As a stable external clock is required for implementing distributed beamforming and MIMO communications, we resorted to attaching each USRP to a stable external clock source.

\subsection{Implementation Details of intra-group synchronization}
The proposed algorithm for timing and frequency synchronization was implemented using GNU Radio. GNU Radio is an open-source toolkit for software defined radios and signal processing. It was used to operate all the USRPs in the experiment. 

We start by describing the parameters of the implemented system and then briefly discuss the details of the implementation. Ten repetitions ($M=10$) of a ZC sequence of length $N_{\text{zc}}=63$ were used in the pilot signal sent by the master.  After transmitting the pilots, 4 ms of guard time was provided for slaves to do their processing and then data was transmitted. The slaves always correlate with the ZC sequence looking for a peak. Once they detect it, they schedule their data transmission after the guard time which is used to estimate the frequency offset by Kalman filtering for smoothing as described earlier. 

 The 4 ms guard time was chosen  as a conservative value since the processing occurs in the PC and the latency needs to be taken into account. One way to reduce such overhead is to implement the synchronization protocol in the FPGA of the USRP, which would would enable us to use a much shorter guard time as was done in \cite{Qiu_2016_MobiHoc}.

As the generation of the master signals and capturing the data for analysis at the receiver nodes is straight forward, we focus on describing the implementation of the slaves. The block diagram of the system is shown in Figure~\ref{fig:USRP_setup}. The input signal from the USRP is sent to the input of an FFT filter which continuously correlates with a ZC sequence. The output of the correlator is passed through a peak detector, which works by taking a moving average and outputing a trigger when the input surpasses ten times the average. This is used to estimate the integer delay $d^\star$ similar to what is described in (\ref{eq:integer_delay_estimator}). This trigger is passed to the \textit{CFO\_EKF} block, which is a custom block written in C++. This block estimates the phase between successive repetitions of ZC sequences and processes the input as described earlier using a Kalman filter to obtain  estimates of the residual frequency error. 
As for calculating the correct time of the transmission and compensating for the frequency and phase errors, we developed the \textit{frac\_timing\_freq\_phase\_synch} block. It takes as input the first trigger corresponding to the first occurrence of a ZC sequence. To calculate the fractional delay $\zeta^{\star}$, this block performs the  processing on the input from the USRP described by (\ref{eq:fractional_delay_estimator}) using  $N_{\zeta} =16$.   Once this trigger is detected and the fractional part is estimated, a burst of data is scheduled to be transmitted after a period equal to the guard time. A frequency synthesizer in this block uses the phase and frequency obtained from the \textit{CFO\_EKF} block to compensate for the error on the data to be transmitted, which is obtained from a vector source. The data after frequency and phase adjustment is then marked at the beginning and the end by burst tags and a tx\_time tag containing the correct transmission time is placed in the first byte of the burst. Burst tags are a feature of the USRP Hardware driver (UHD), which allows the USRP to perform bursty transmissions. The tx\_time tag allows users to have a high accuracy control of the transmission  time of a data burst.  It works by sending a timestamp of the desired transmission time to the USRP along with the data. The USRP delays the burst transmission until the correct time has come. This timestamp is not transmitted over the air; it is used only to control the USRP.   

As mentioned in Section~\ref{sec:algorithm}, our timing synchronization algorithm does not distinguish between the hardware timing offset and propagation delay and therefore the accuracy highly depends on differences in intra-group distances among radios. In our experiment, the relative distances among the transmit nodes were less than 1m, which results in accuracy biasing up to 3.3ns. The distance between the transmitter and the slaves was less than 3m and the SNR was controlled by adjusting the transmission power gain in the USRPs. A synchronization protocol that incorporates propagation delay is left for future work as this delay becomes more critical when intra-group distance and signal bandwidth become larger.

\section{Experiment Results}
\label{sec:results}
In this section, we present the experimental synchronization accuracy using the SDR platform.

\subsection{Experimental Setup}
\begin{figure}
\centering
\includegraphics[width=3in]{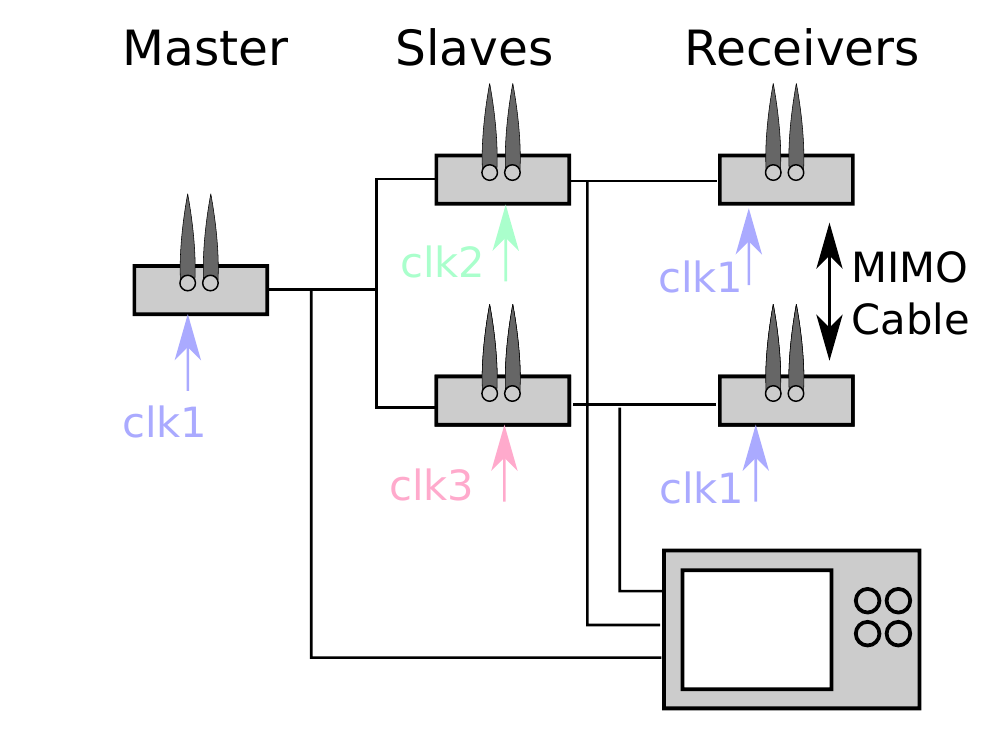}
\caption{\bf{Experiment setup for residual CFO and STO measurement.}}
\label{fig:experiment_setup1}
\end{figure}

\begin{figure}
\centering
\includegraphics[width=3in]{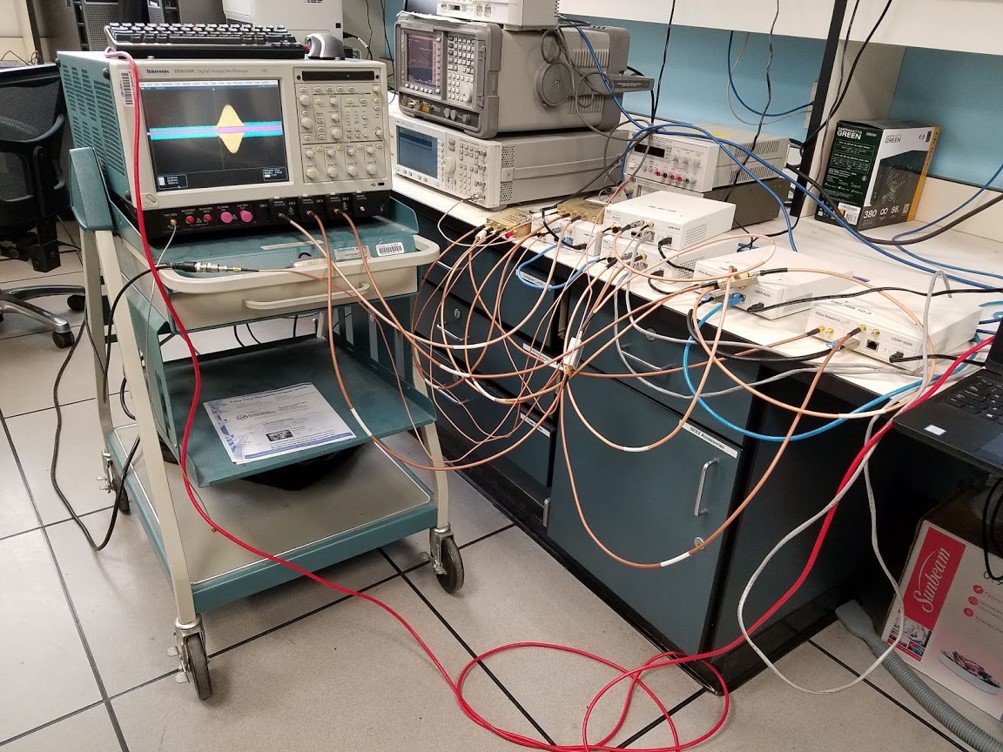}
\caption{\bf{Experiment setup for residual CFO and STO measurement.}}
\label{fig:experiment_setup2}
\end{figure}

 Our proposed protocol and algorithms were tested using a system consisting of one master and two slaves. A high sampling rate oscilloscope and two additional USRPs were used to evaluate the timing and frequency synchronization. To overcome the instability of the internal oscillator of the USRP, each of the master and two slaves was connected to a unique external oscillator. The two USRPs used for evaluating the system were synchronized using a MIMO cable to avoid having any timing and frequency drifts between them to provide accurate measurements of the performance of the slaves. The oscilloscope used was the  Tektronix DLS6154, which has a sampling rate up to 40 GSample/s, which enabled us to obtain a high resolution estimate of any remaining timing error between the master and slaves. A schematic of this setup is shown in Figure~\ref{fig:experiment_setup1}, while a photo of the actual setup is shown in Figure~\ref{fig:experiment_setup2}.

 For this experiment, the carrier frequency used  was $f_\text{c}=2.4$~GHz  and the signal bandwidth of the master and slaves was $1/\Ts=1$ MHz. The two synchronized USRPs were used to measure the residual CFO (RCFO). Because the RCFO was below 50 Hz, the measuring USRPs used a lower sampling rate of 0.25MS/s to capture the compensated signals from the slaves. Then the captured signal was digitally downsampled by a factor of 250. By using a 256-point FFT for our RCFO evaluation, we were able to measure the RCFO with a resolution of 0.5Hz.
For measuring the residual timing offset we used the Tektronix DLS6154 oscilloscope.  Data was recorded using the oscilloscope with a time resolution of 800ps. The residual timing error was estimated by correlating the captured waveform from the master and one of the slave USRPs. A total of  650 waveforms were captured during a period of four hours in order to incorporate long-term  effects such as heating.
 


The host computers used as slaves were a Dell precision 3520 having a Xeon E3-1505M V6 processor and 8 GB of RAM, and a Thinkpad T430 having  an Intel I7-3520M processor with 8 GB of RAM. The master and the evaluation USRPs used a Thinkpad with similar specs as host.
We start by showing the residual frequency offset results followed by the residual timing offset results.


\subsection{Carrier Synchronization Results}
\begin{figure}
\centering
\includegraphics[width=3.4in]{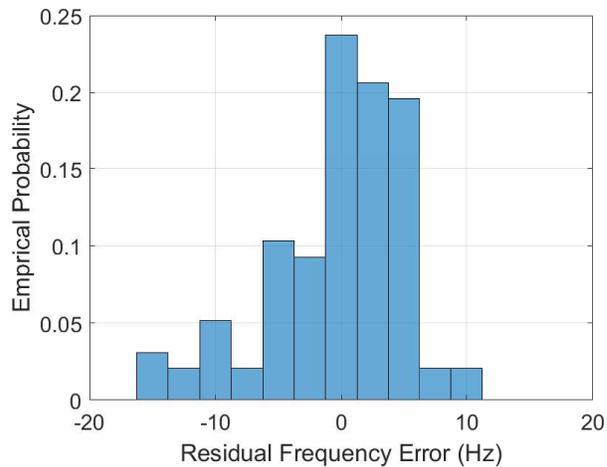}
\caption{\bf{Histogram of measured residual carrier frequency synchronization error.}}
\label{fig:freq_sync_CDF}
\end{figure}

The residual CFO  between the two slaves was calculated  based on measurements captured from the two slaves using the evaluation USRPs. Figure~\ref{fig:freq_sync_CDF} shows the histogram of the measured RCFO with each bin having a width of $2.5$ Hz.  The mean, standard deviation, and tail probability of the residual error
are summarized in Table~\ref{tbl:freq}. Due to the fact that  phase synchronization, i.e., intra-group channel estimation, is not implemented, we use the results of  \cite{freq_benchmark} as a benchmark where the mean residual frequency error is provided in the table.

\subsection{Timing Synchronization Results}
\begin{figure}
\centering
\includegraphics[width=3.4in]{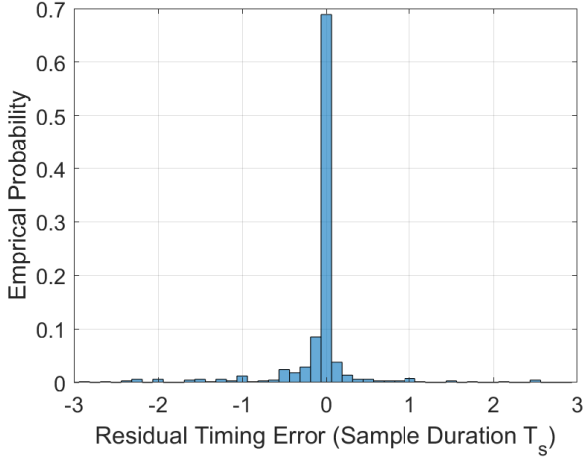}
\caption{\bf{Histogram of measured residual sample timing synchronization error.}}
\label{fig:timing_sync_CDF}
\end{figure}

 Figure~\ref{fig:timing_sync_CDF} shows the histogram of the  residual timing offset between the master and one of the slaves. Each bin has a width of $\Ts/8$, meaning the center bin represents residual error within $\pm \Ts/16$. The mean, standard deviation and tail probability of the residual error in terms of sample duration $\Ts$ are summarized in Table~\ref{tbl:time}. A comparison with benchmark results from \cite{Qiu_2016_MobiHoc} and \cite{Brown_2017_ICASSP} is also included. Note that \cite{Qiu_2016_MobiHoc} does not intend to adjust CFO and the STO synchronization accuracy can be affected by an actual multipath environment due to its fully wireless setting. Works \cite{Brown_2017_ICASSP} and \cite{Brown_2018_aeroconf} are tested in a wired environment where the carrier is perfectly synchronized among radios. 


\begin{table}
\caption{Measured Residual Frequency Error Statistics and Comparison with Benchmarks}
\label{tbl:freq}
\centering
\def\arraystretch{1.5}
\begin{tabular}{l|c|r}
\hline 
\textbf{Item} & \textbf{Expression} & \textbf{Value}\tabularnewline
\hline 
\multicolumn{3}{c}{This work}\tabularnewline
\hline
Mean [Hz]& $\mu(\Delta_{\text{R,F}})$& -0.322 \tabularnewline
Std. Dev. [Hz]& $\sigma(\Delta_{\text{R,F}})$ & 5.253 \tabularnewline
\hline
Tail prob. at 5 Hz & $\text{prob}(|\Delta_{\text{R,F}}|> 5)$ &  0.247 \tabularnewline
Tail prob. at 10 Hz & $\text{prob}(|\Delta_{\text{R,F}}|> 10)$ &  0.072 \tabularnewline
Tail prob. at 15 Hz & $\text{prob}(|\Delta_{\text{R,F}}|> 15)$ &  0.000 \tabularnewline
\hline 
\multicolumn{3}{c}{Globecom'12 \cite{freq_benchmark}}\tabularnewline
\hline 
Mean [Hz]& $\mu(\Delta_{\text{R,F}})$ & $<$0.600 \tabularnewline
\hline 
\end{tabular}
\end{table}

\begin{table}
	
	\caption{Measured Residual Timing Error Statistics and Comparison with Benchmarks}
	\label{tbl:time}
	\centering
	\def\arraystretch{1.5}
	\begin{tabular}{l|c|r}
		\hline 
		\textbf{Item} & \textbf{Expression} & \textbf{Value}\tabularnewline
		\hline 
		\multicolumn{3}{c}{This work  (wired, $1/\Ts=1$ MHz)}\tabularnewline
		\hline
		Mean [$\Ts$] & $\mu(\epsilon_{\text{R,T}})$ & 0.084 \tabularnewline
		Std. Dev. [$\Ts$] & $\sigma(\epsilon_{\text{R,T}})$ & 0.516\tabularnewline
		\hline 
		Tail prob. at $\Ts/16$ & $\text{prob}(|\epsilon_{\text{R,T}}|> 1/16)$ &  0.311\tabularnewline
		Tail prob. at $\Ts/8$ & $\text{prob}(|\epsilon_{\text{R,T}}|> 1/8)$ &  0.310\tabularnewline
		Tail prob. at $\Ts/4$ & $\text{prob}(|\epsilon_{\text{R,T}}|> 1/4)$ &  0.161 \tabularnewline
		Tail prob. at $\Ts/2$ & $\text{prob}(|\epsilon_{\text{R,T}}|> 1/2)$ &  0.109 \tabularnewline
		Tail prob. at $\Ts$ & $\text{prob}(|\epsilon_{\text{R,T}}|> 1)$ &  0.059 \tabularnewline
		Tail prob. at $2\Ts$ & $\text{prob}(|\epsilon_{\text{R,T}}|> 2)$ &  0.023 \tabularnewline
		\hline 
		\multicolumn{3}{c}{MobiHoc'16 \cite{Qiu_2016_MobiHoc} (wireless, $1/\Ts = 20$ MHz)}\tabularnewline
		\hline 
		Mean [$\Ts$] & $\mu(\Delta_{\text{R,F}})$ & 1.874 \tabularnewline
		Std. Dev. [$\Ts$]& $\sigma(\Delta_{\text{R,F}})$ & 1.938 \tabularnewline
		\hline 
		\multicolumn{3}{c}{ICASSP'17 \cite{Brown_2017_ICASSP} (wired, $1/\Ts = 150$ KHz)}\tabularnewline
		\hline 
		Mean$^{a}$ [$\Ts$] & $\mu(\Delta_{\text{R,F}})$ & 0.541 \tabularnewline
		Std. Dev. [$\Ts$]& $\sigma(\Delta_{\text{R,F}})$ & 0.020 \tabularnewline
		\hline 
		\multicolumn{3}{c}{AeroConf'18 \cite{Brown_2018_aeroconf} (wired, $1/\Ts = 250$ KHz)}\tabularnewline
		\hline 
		Mean [$\Ts$] & $\mu(\Delta_{\text{R,F}})$ & 0.000 \tabularnewline
		Std. Dev. [$\Ts$]& $\sigma(\Delta_{\text{R,F}})$ & 0.001 \tabularnewline
		\hline 
		\multicolumn{3}{l}{a. Empirical calibration further improves accuracy.}\tabularnewline

	\end{tabular}
\end{table}

\section{Discussion}
\label{sec:discussion}
The performance results for timing and frequency synchronization show that the synchronization system achieved the requirements for a  typical distributed beamforming scenario with the parameters described in Section \ref{sec:results}.  These results are comparable to performance that can be obtained using a GPSDO as  was mentioned in Section \ref{sec:implementation}, without the added cost of the GPSDO module or the limitations of GPS signals.  By using an EKF over a periodic preamble, we get higher accuracy than what is achievable without the EKF as calculated from the CRLB (\ref{eq:CFO_CRLB}).  The achievable accuracy in frequency estimation using the EKF depends on  the quality of the reference clock in terms of frequency drift and phase noise.
So far, the current experiments used  external independent high-end reference clocks from measurement equipment. The use of measurements with more compact and portable reference clocks is of particular interest for future work. 
As for the residual timing error results, although 70\% of the measurements are within $\pm T_{s}/16$,  we have observed  long tails in the empirical error distribution. This deteriorates the standard deviation of the results and makes it far from what the CRLB (\ref{eq:STO_CRLB}) predicts. Further investigation is required to diagnose this problem and solve.
The current model assumes a narrowband system and experiments were conducted in an over-the-wire setting where there is no multipath environment. Development of the synchronization to support wideband systems is of high importance. On one hand, the synchronization requirements would vary drastically if we used a cyclic-prefix based waveform. These requirements need to be analyzed for the exact type  of waveform. We are particularly interested in   Orthogonal Frequency Division Multiplexing (OFDM), Single-Carrier Frequency Domain Equalization (SC-FDE) which features low peak to average power ratios, and  Orthogonal Time Frequency and Space (OTFS) which is resilient in high-mobility environments. On the other hand,  the synchronization protocol should be adapted to retain robustness and ahieve the requirements of a specific waveform in a multipath environment.
One possible way to extend this work is by porting the implementation from the host PC to the FPGA. This will enable us  to reduce the required latency and to deploy the system using higher sample rates.

\section{Conclusion}
\label{sec:conclusion}
In this work, we have developed and implemented a joint carrier frequency and sample timing synchronization algorithm and protocol for cooperative communication in distributed arrays. Using the received baseband signal and proposed digital signal processing techniques, different radios with separate reference clocks can achieve less than 5 Hz residual frequency offset with 75 percent. The residual timing offset is within $1/16$ of symbol duration of preamble by 68 percent. Our simulations show that this specification provides near optimal signal power gains when distributed beamforming is used for range extension applications in a typical setting.

\appendices{}              
%
%
\section{Nomenclature}
\label{app:notation}

Notations in the main context are summarized in the Table~2. The value of parameters used in USRP implementation, when applicable, is also provided.
\def\arraystretch{1.5}
\begin{table}
\caption{Nomenclature}
\centering
\begin{tabular}{c|c|c}
\hline 
Symbol & Value &Explainations\tabularnewline
\hline 
\hline 
$\Nt$, $\Nr$ & - & Num. of radio in T/Rx group\tabularnewline
\hline 
$n$ & - & Time index of discrete-time signal\tabularnewline
$m,M$ & 10 & Index and number of ZC repetions\tabularnewline
$N_{\text{zc}}$ & 63 & Length of ZC sequence \tabularnewline
$N_{\zeta}$ & 16 & Frac. delayed candidates set size. \tabularnewline
$\mathcal{Z}$ & - & Frac. delay candidates set. \tabularnewline
$h_0$ & - & Channel between mater and slave \tabularnewline
$\sigma_{\text{n}}^{2}$ & - & AWGN power in sync. \tabularnewline
$\text{SNR}_{\text{sync}}$ & - & SNR in intra-group sync. \tabularnewline
\hline 
$\Ts$ & 1 $\mu$s & Sample duration\tabularnewline 
$T_{\text{guard}}$ & 4 ms & Period of SS bursts\tabularnewline
$T_{\text{CYC}}$ & - & Period of SS bursts\tabularnewline
\hline 
$\Delta f,\Delta \hat{f}$ & - & CFO and estimate in [Hz] \tabularnewline
$\eF,\hat{\epsilon}_{\text{F}}$ & - & Normalized CFO and estimate in [rad]\tabularnewline
$\Delta_{\text{R,F}}$ & - & Post-comp. residual CFO in [Hz]\tabularnewline
\hline
$\Delta \tau$ & - & TO and estimate in [sec] \tabularnewline
$\eT,\hat{\epsilon}_{\text{T}}$ & - & Normalized TO and estimate in [$\Ts$] \tabularnewline
$\epsilon_{\text{R,T}}$ & - & Post-comp. residual STO in [$\Ts$]\tabularnewline
$d^{\star},\hat{d}^{\star}$ & - & Int. TO in [$\Ts$] and estimates\tabularnewline
$\zeta^{\star},\hat{\zeta}^{\star}$ & - & Frac. TO in [$\Ts$] and estimates\tabularnewline
\hline
\multicolumn{3}{c}{Signal and Sequences}\tabularnewline
\hline
$s[n]$ & - & Transmit sync. preamble \tabularnewline
$s_{\text{zc}}[n]$ & - & ZC sequence w/ finite t-support $N_{\text{zc}}$\tabularnewline
$r[n]$ & - & Received sync. preamble \tabularnewline
$y_{\text{corr}}[n]$ & - & Received signal after ZC correlation\tabularnewline
$\tilde{r}[n]$ & - & Received post-CFO-comp. signal\tabularnewline
$\tilde{s}^{(\zeta)}_{\text{zc}}[n]$ & - & $\tau$-fractional delayed ZC sequence \tabularnewline
$p_{\text{ps}}(t)$ & - & RF pulse shaping function \tabularnewline
\hline 
\end{tabular}
\end{table}

\acknowledgments
This work was supported in part by the Defense Advanced Research Projects Agency (DARPA) under contract D17PC00006. The views, opinions and/or findings expressed are those of the author and should not be interpreted as representing the official views or policies of the Department of Defense or the U.S. Government.

This work was supported in part by the CONIX Research Center, one of six centers in JUMP, a Semiconductor Research Corporation (SRC) program sponsored by DARPA.

\bibliographystyle{IEEEtran}
\bibliography{references,IEEEabrv}

\end{document}